\documentclass[preprint,3p,12pt,numbers,sort&compress]{elsarticle}
\usepackage{lipsum}
\usepackage{nomencl}
\makenomenclature
\makeatletter
\def\ps@pprintTitle{%
 \let\@oddhead\@empty
 \let\@evenhead\@empty
 \def\@oddfoot{}%
 \let\@evenfoot\@oddfoot}
\makeatother
\usepackage{array}
\usepackage{booktabs}
\DeclareUnicodeCharacter{2212}{-}
\DeclareUnicodeCharacter{00A0}{ }
\usepackage{miller}
\usepackage{graphicx,float}
\usepackage[colorlinks=true, allcolors=blue]{hyperref}
\usepackage{subcaption}
\usepackage{varioref}
\usepackage{ulem}
\usepackage{amsmath,amsthm,amssymb,amsfonts}
\usepackage[linesnumbered,ruled,vlined]{algorithm2e}
\RestyleAlgo{ruled}
\usepackage[noend]{algpseudocode}
\usepackage{cases}
\usepackage{tabularx}
\usepackage{makecell}
\usepackage[capitalise]{cleveref}
\usepackage{array}
\usepackage{tikz}
\usepackage{makecell}
\usepackage{booktabs,tabularx}
\usepackage{mwe}
\usepackage{xfrac}
\usepackage{mathtools}
\usepackage{epstopdf}
\biboptions{sort&compress}
\usepackage{bm,upgreek}

\usepackage{color}
\usepackage[bordercolor=white,backgroundcolor=gray!30,linecolor=black,colorinlistoftodos]{todonotes}

\usepackage[section]{placeins}

\usepackage{amsmath}
\makeatletter
\@namedef{ver@amsmath.sty}{}
\makeatother
\usepackage{amstext}
\usepackage{soul}
\usepackage{xcolor}

\newcommand{\alr}[1]{\textcolor{red}{#1}}

\begin{document}
\begin{frontmatter}

\title{Towards Development of Automated Knowledge Maps and Databases for Materials Engineering using Large Language Models}

\author[affil1]{Deepak Prasad \corref{cor1}}
\address[affil1]{Artificial Intelligence and Data Science, Vivekanand Education Society Institute of Technology, Mumbai, India}
\author[affil1]{Mayur Pimpude \corref{cor2}}
\author[IITB,cminds]{Alankar Alankar\corref{cor3}}
\address[IITB]{Department of Mechanical Engineering, Indian Institute of Technology Bombay, Mumbai 400076, India}
\address[cminds]{Center for Machine Intelligence and Data Science, Indian Institute of Technology Bombay, Mumbai 400076, India}
\ead{alankar.alankar@iitb.ac.in}
\cortext[cor3]{Author. Tel.: +91-9769415356, Fax: +91-22-25726875}

\begin{abstract}
In this work a Large Language Model (LLM) based workflow is presented that utilizes OpenAI ChatGPT model GPT-3.5-turbo-1106 and Google Gemini Pro model to create summary of text, data and images from research articles. It is demonstrated that by using a series of processing, the key information can be arranged in tabular form and knowledge graphs to capture underlying concepts. Our method offers efficiency and comprehension, enabling researchers to extract insights more effectively. Evaluation based on a diverse Scientific Paper Collection demonstrates our approach in facilitating discovery of knowledge. This work contributes to accelerated material design by smart literature review. The method has been tested based on various qualitative and quantitative measures of gathered information. The ChatGPT model achieved an F1 score of 0.40 for an exact match (ROUGE-1, ROUGE-2) but an impressive 0.479 for a relaxed match (ROUGE-L ,ROUGE-Lsum) structural data format in performance evaluation. The Google Gemini Pro outperforms ChatGPT with an F1 score of 0.50 for an exact match and 0.63 for a relaxed match. This method facilitates high--throughput development of a database relevant to materials informatics. For demonstration, an example of data extraction and knowledge graph formation based on a manuscript about a titanium alloy is discussed.
\end{abstract}


\end{frontmatter}

\section{Introduction}
\label{introduction}
\noindent Machine learning has proved to be an efficient tool for understanding correlations in data, mechanisms, predicting next experiments and physics based computational model for new material discovery. For high-throughput design of technologically premium materials e.g. battery materials, hydrogen storage materials, and aerospace materials, a balanced combination of computational modeling and synthesis is desired. A majority of machine learning models are data--driven models that loose applicability in the absence of versatile and reliable data. A large amount of rigorously validated data is available in the form of peer-reviewed journals including numerical data, graphics, and text-based logical reasoning \& facts. Although, there has been an on--going effort to collect numerical data and context learning \cite{mikolov2013efficient} using natural language processing (NLP) applied to literature \cite{mausambert2022}, the collection of graphical data and driving context thereof is not routinely done. A detailed review of language models for materials science and their evolution is presented in \cite{hubueehler2023}. Efficiently analyzing research papers has been a long-standing challenge within the academic community.
\par Traditional methods often require researchers to invest substantial time in reading and deciphering complex papers. Although being robust, this method may not be the most suited for high--throughput data collection. However, recent advancements in natural language processing (NLP) \cite{nlpbooklee2024, eisenstein2019introduction} have opened up new avenues for automating this process. A significant fraction of such \textit{language} models are based on Bi--directional Encoder Representations from Transformers (BERT) \cite{devlin-etal-2019-bert}. BERT based models offer significant advancement as extraction tools for a general corpus of science. Such models are not trained specifically on Materials Science jargon and may not capture context. Also, noteworthy is that most such models have demonstrated the performance based on classification of text alone. Further, larger fraction of such text is a corpus of abstracts e.g \cite{beltagy2019scibert, mausambert2022}. A more application orientated approach could be to collect operable data that can be further used for analyses, either directly or in the existing physics driven numerical models.

\par Some of the earliest applications of NLP in materials science are seen in the work of Kim et al \cite{kim2017}, Jensen et al \cite{jensen2019} who used NLP for automated data extraction for accelerated synthesis of metal oxide and zeolites respectively. Shetty and Ramprasad \cite{SHETTY2021101922} applied NLP for automated extraction of knowledge on a corpus of 0.5 M journal articles and were able to predict novel polymers based on the text-based approach solely. Following SciBERT \cite{beltagy2019scibert}, Gupta et al, developed a MatSciBERT model \cite{mausambert2022}. ChatGPT has been used for assisting with additive manufacturing \cite{BADINI2023278} and to extract data from research papers using prompt engineering \cite{polak2023extracting}, amongst many other routine applications e.g. write computer programs etc.

\par Overall, \textit{named entity recognition} (NER) techniques have played a pivotal role in the aforementioned realm of extracting structured information from unstructured data. Consideration of parts of a text as a named entity (NE) depends on the application that makes use of the annotations. One such application is document retrieval or automated document forwarding in which documents annotated with NE information can be searched more accurately than raw text \cite{mikheev-etal-1999-named}. However, it often struggles to provide concise and structured outputs, particularly when dealing with individual paragraphs. A more comprehensive understanding is achieved using LLMs \cite{ChatGPT2019}. While LLMs mark a significant stride in NLP by capturing intricate relationships between words and entities, they primarily focus on relations between individual words, lacking a readily comprehensible structural format for unstructured text. LLMs open exciting opportunities for extracting insights from text at large scale \cite{oppenlaender2023mapping}.

\par In practice, the valuable relations extracted by NER and LLMs may not inherently provide an easily readable or recognizable structure for unstructured text. Prompt engineering is the means by which LLMs are programmed via prompts \cite{white2023prompt}. The challenge lies in presenting this information in a format that enhances both comprehension and analysis. Knowledge Graphs are structured representations of information that capture the relationships between entities in a particular domain \cite{trajanoska2023enhancing}. Different approaches have emerged including those based on NER and ontology--based methods that are domain--specific. Additionally, LLM models like OpenAI’s ChatGPT \cite{chatgpt}, Google Gemini Pro have been utilised to extract key value pairs i.e. that data that are linked with each other. While these methods contribute to knowledge mapping, the resulting graphs can be dense and challenging to interpret.
Our primary objective is to establish a work flow using LLMs models like OpenAI ChatGPT (GPT-3.5-turbo-1106) and Google Gemini Pro for creating knowledge maps (graphs) that identifies important process steps for materials, valuable relationships, important property data and lists salient outcomes of research articles. This approach aims to develop a high--throughput, strong context--based material database extracted from available open literature.

\section{Methodology}\label{sec:methodology}



\begin{figure}[!h]
  \centering
  \includegraphics[width=0.7\textwidth]{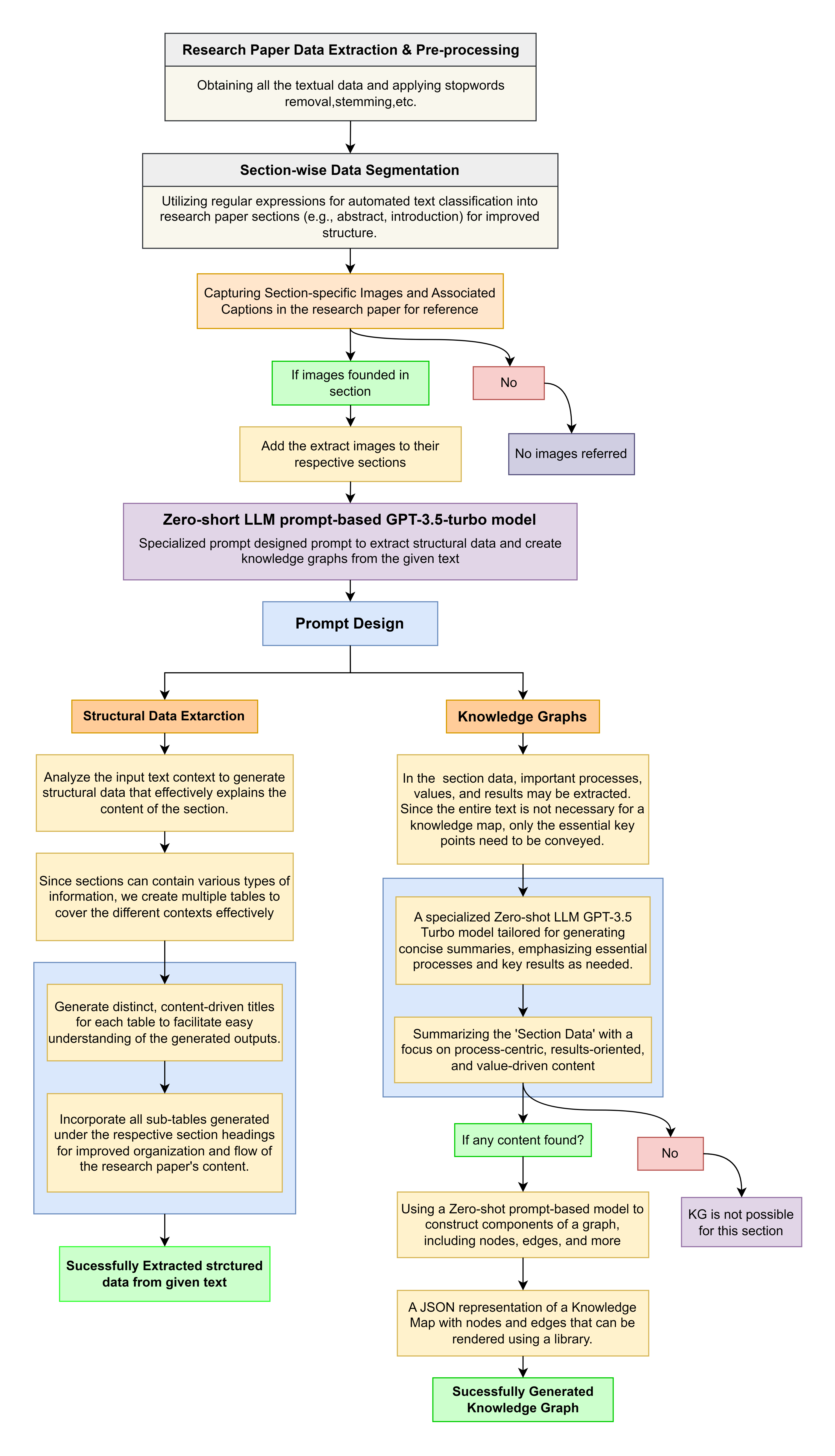}
  \caption{Flowchart for data extraction and formation of knowledge graph assisted by ChatGPT and Gemini Pro.}
  \label{fig:overallflow}
\end{figure}

\noindent In this work, the collected text from the journal articles has been used to extract knowledge graphs, construct data tables. Overall, we have exercised extraction of 100 relations from seven peer-reviewed journal articles for demonstration. A novel text and image extraction approach allows to systematically capture section--specific text and images from the portable document format (pdf) files of journal articles. The quality assurance measures employed in this work include cross-referencing the extracted content with the original papers to ensure accuracy. The meticulously collected text has served as the foundation of our work. Overall workflow of the methodology is shown in Figure \ref{fig:overallflow}.

\par
\begin{figure}[!h]
  \centering
  \includegraphics[width=\textwidth]{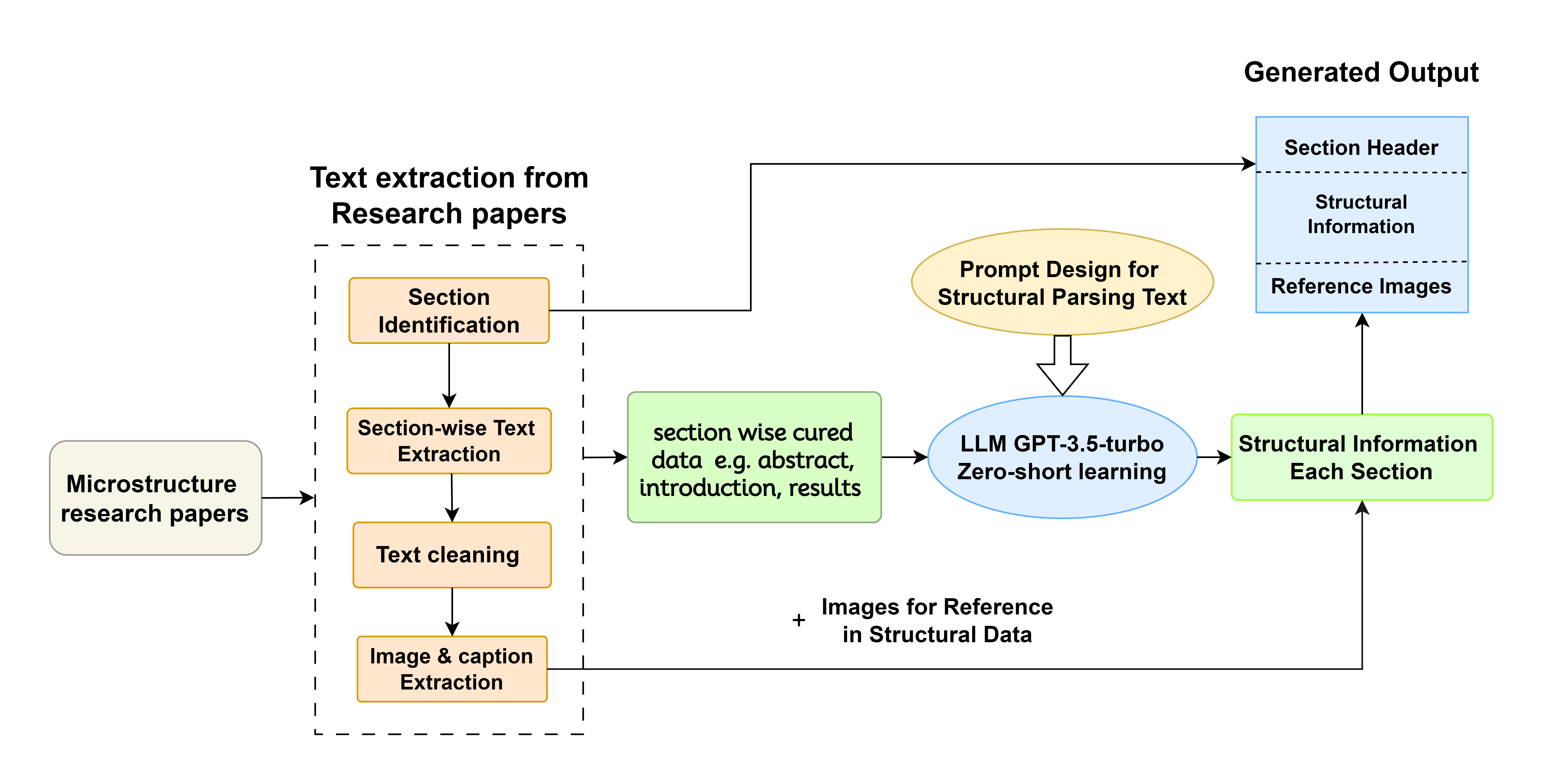}
  \caption{Flow of structural text extraction using prompt engineering.}
  \label{fig:structural_text}
\end{figure}

 In the initial phase of our data collection approach, advanced techniques are employed such as tokenization and stemming to extract significant textual insights from research papers. This process involves breaking down the text into tokens and reducing words to their root forms, enhancing our ability to identify key entities and relationships. The extracted content is split into sections as defined in the journal article, allowing for focused analyses within specific segments of the papers. This meticulous extraction and organization process helps the model understand specific context of the segment of the scientific article to generate more accurate summary and also to easily interpret the output. This concept of structural text extraction using prompt engineering is shown in Figure \ref{fig:structural_text}.

\subsection{Extraction of structured information from text}

\begin{figure}
  \centering
  \includegraphics[width=\textwidth,trim={0 0.0cm 0 0.0},clip]{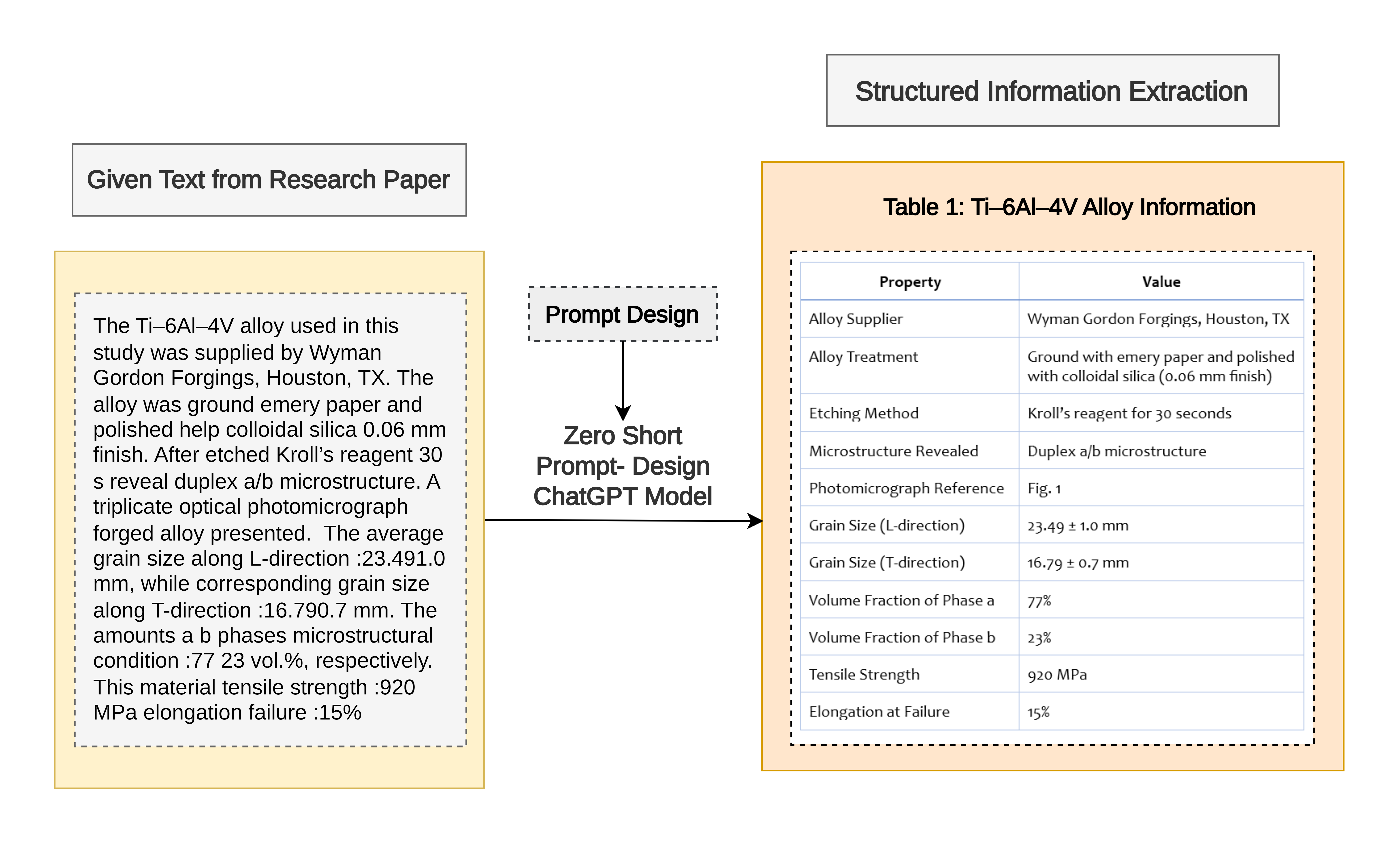}
  \caption{Generation of structural formats from research paper text using ChatGPT. The above demonstration is for the research article by Tarzimoghadam et al. \cite{TARZIMOGHADAM2015291}.}
  \label{fig:structural_format_chatgpt}
\end{figure}

\noindent To extract structural data and numerical values from material science research papers, specialized prompts are utilized that are designed to guide prompt-based LLMs models utilizing ChatGPT and Gemini models. These prompts are tailored to capture relevant material properties, experimental conditions, and performance metrics. By employing these targeted prompts, it is ensured that accurate structural data and numerical values are extracted enabling the generation of comprehensive and precise representations of the content of research papers. See Figure \ref{fig:structural_format_chatgpt}.

The structured responses generated by the zero-shot LLMs models are seamlessly merged with the associated images that are referenced within the employed structured format. This integration involves combining the model's textual output with the referenced images, resulting in a cohesive representation that enriches the overall comprehension of the content. This interconnected approach ensures that textual and visual components are presented together.

\subsection{Knowledge Graph Generation}

\begin{figure}
  \centering
  \includegraphics[width=1.0\linewidth,trim={0 0.0cm 0 0.0},clip]{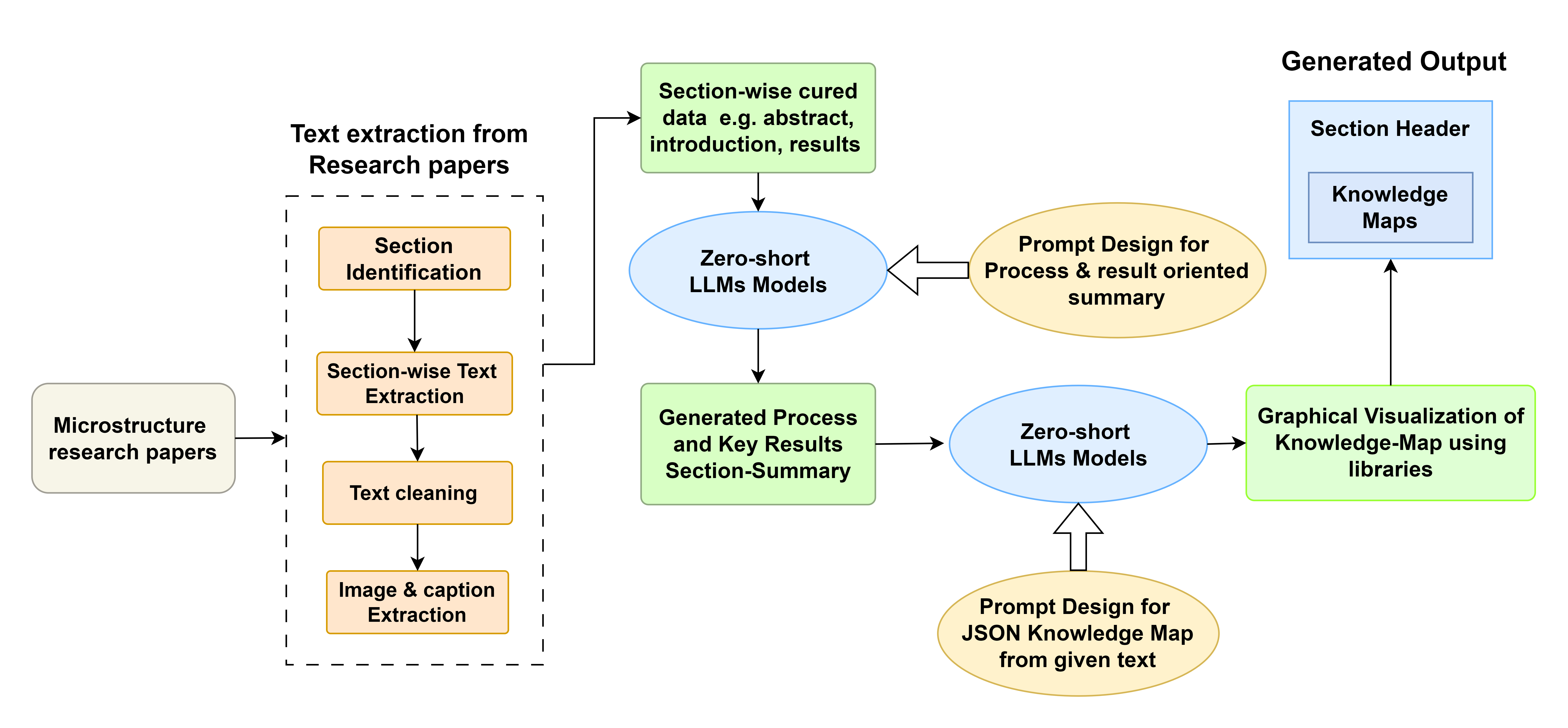}
  \caption{Workflow for Knowledge Graph generation from text data.}
  \label{fig:knowledge_graph_concept}
\end{figure}

\noindent A knowledge graph is a structured representation that visually illustrates relationships between concepts or entities, facilitating the understanding of complex information domains through organized visualization. This method utilizes nodes and edges to depict interconnected information. The process involves extracting essential information and significant data from text using these models. See Figure \ref{fig:knowledge_graph_concept}. It is then converted into JavaScript Object Notation (JSON) format \cite{Crockford2006TheAM}. JSON is a lightweight data interchange format used for transmitting data between servers and web applications. Its human-readable nature and ease of comprehension by both humans and machines make it widely used.

Moreover, JSON's simplicity and flexibility allow for the easy representation and storage of the knowledge graph. Knowledge graphs function as repositories of organized knowledge, represented as a set of triplets $KG = (h, r, t)$ in $E \times R \times E$ format, where $E$ represents sets of entities and $R$ represents sets of relations. JSON's straightforward structure accommodates the convenient storage of node attributes, edges, and their interconnections within a knowledge graph. This simplicity facilitates efficient handling and retrieval of graph--related data, enhancing ease in managing complex relationships between entities. The use of JSON thereby enables the storage and retrieval of this structured data, facilitating efficient manipulation and utilization of knowledge graphs across various applications.

\subsection{Relation-Extraction Methods}
\noindent Relation extraction is a fundamental NLP task that aims to identify semantic linkages between entities within sentences or documents. While capturing relationships within the material science field might seem less significant, the emphasis lies in understanding the involved processes and extracted results crucial for generating a comprehensive knowledge map. Such knowledge maps are eventually useful for development of new material and corresponding process optimization. A comparative study of entity relationship extraction using ChatGPT is discussed in the following.

\subsubsection{LLMs--based Relation Extraction}\label{subsec:relationship_extraction}

\noindent In this method LLM models OpenAI’s ChatGPT and Google Gemini Pro are utilized. It is essential to note that this constitutes one of the steps of creating a knowledge graph, not the final iteration in the overall workflow that we propose in the current work. The novelty of our work lies in devising a specific prompt to identify the desired linkages using LLMs. These linkages are stored in JSON format serving as the foundation for our knowledge graph.
\par Despite LLMs capability to identify items and establish linkages, it encounters difficulty in comprehending the primary materials related processes and significant scientific details from the provided text. Even with the incorporation of a material science tokenization, it is unable to extract key process details and results adequately. It succeeds in linking words but is not able to grasp the sequence of events and outcomes. For instance, in our exploration of titanium-based alloys, we employ LLM Models GPT-3.5-turbo-1106 and Google Gemini Pro to analyze an article focused on an aerospace-grade Ti--based alloy. The article is by authors Tarzimoghadam et al \cite{TARZIMOGHADAM2015291}. LLMs effectively identifies the constituent elements, notably Ti, within the research paper. However, it has faced challenges in comprehending the intricate fabrication methodologies crucial for attaining the alloy's specified strength and high-temperature resilience, as detailed in the work.

\subsubsection{Extracting Relationships: Process-Centric Insights with LLMs} 
\label{subsec:summary_relationship_extraction}
\noindent The second approach involves generating a concise summary of the research paper, emphasizing key constraints, the involved process, and important insights. Despite experimenting with a BERT summarization model, it failed to meet specific constraints related to the summarization process. Consequently, a customized prompt was designed specifically for LLMs to ensure that the generated summary aligns with specified requirements and maintains scientific consistency, as previously mentioned. The encountered issues during the BERT-based summarization attempts, particularly regarding the identified constraints, have prompted efforts to address and rectify the shortcomings without further elaboration on the specific nature of these constraints.

\section{Prompt Engineering}
\noindent Prompt engineering stands as an emerging field dedicated to the deliberate formulation and refinement of prompts with the primary objective of optimizing the efficacy of LLMs across a spectrum of applications and research domains. Within the context of this research, a prompt is construed as a series of natural language inputs designed for specific LLM tasks. This construct comprises three integral components: (1) instruction -- serving as a succinct directive guiding the model's task execution; (2) context -- providing contextual information or few--shot examples to augment the input text; and (3) -- input text, representing the textual data requiring processing by the model.

\par One key method in prompt engineering is the Chain-of-Thought (CoT) approach \cite{Pan_2024}. It's special because it breaks down tricky thinking into smaller, easier steps. This helps reshape how we interact with prompts, making it simpler for the models to understand. Another important component in prompt engineering is bringing in outside information. By adding specific details related to certain topics in the prompts, we give the models more context. This helps the model perform better by having the right information to generate accurate and detailed responses.

\par
\begin{figure}[!h]
  \centering
  \includegraphics[width=\textwidth]{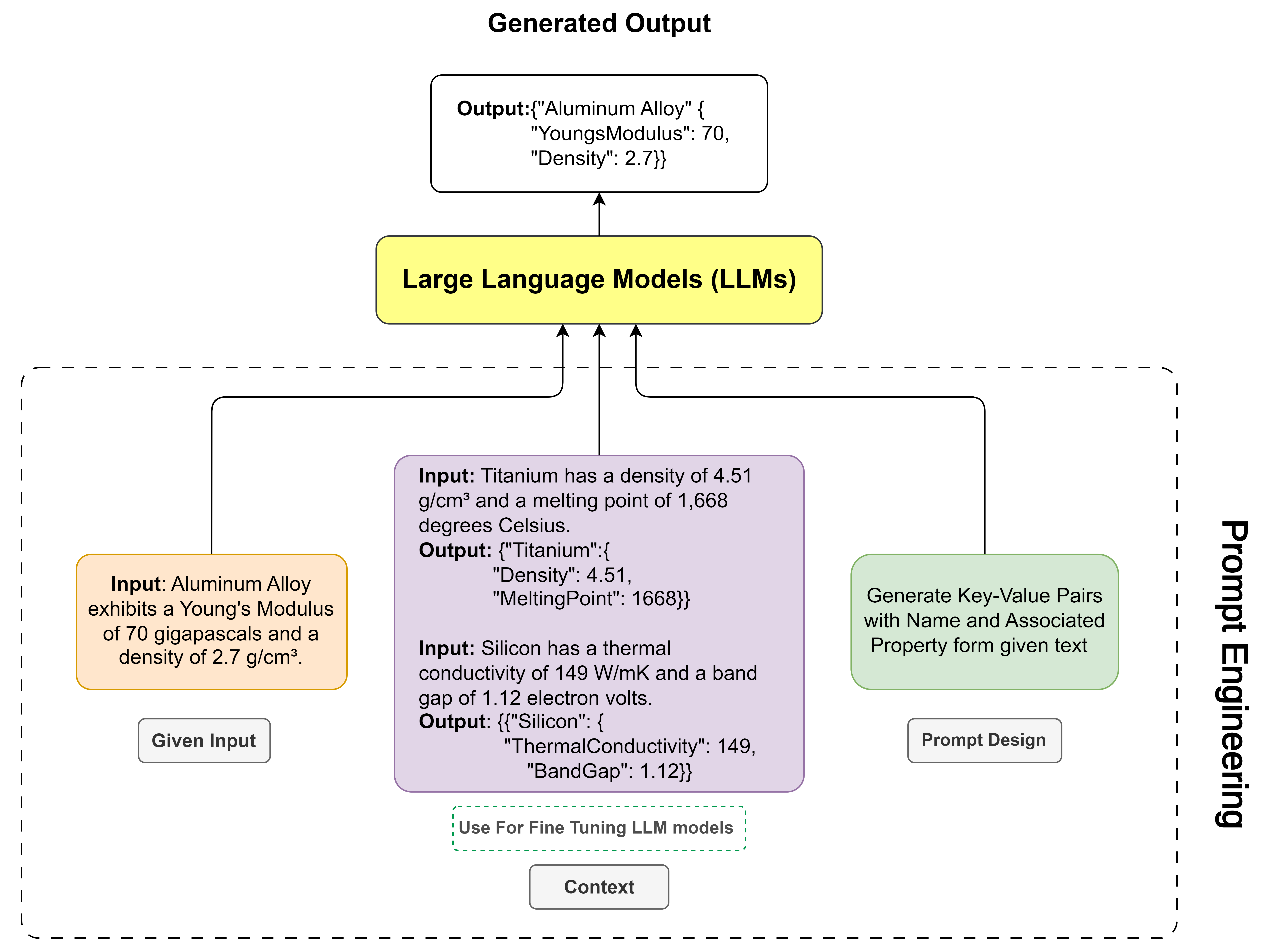}
  \caption{The examples of prompt engineering, one for aluminum and the other for titanium are shown.}
  \label{fig:prompt_engineering_working}
\end{figure}

 To extract structural data and numerical values from Material Science and Engineering research papers, specialized prompts are utilized that are designed to guide prompt-based models. These prompts are tailored to capture relevant material properties, experimental conditions, and performance metrics. Figure \ref{fig:prompt_engineering_working} shows one instance of output generated by prompt operated on the text of the journal paper using ChatGPT. By employing these targeted prompts, it is ensured to extract the accurate structural data and numerical values, enabling the generation of comprehensive and precise representations of the research papers' content.

\par
\begin{figure}[!h]
  \centering
  \includegraphics[width=\textwidth]{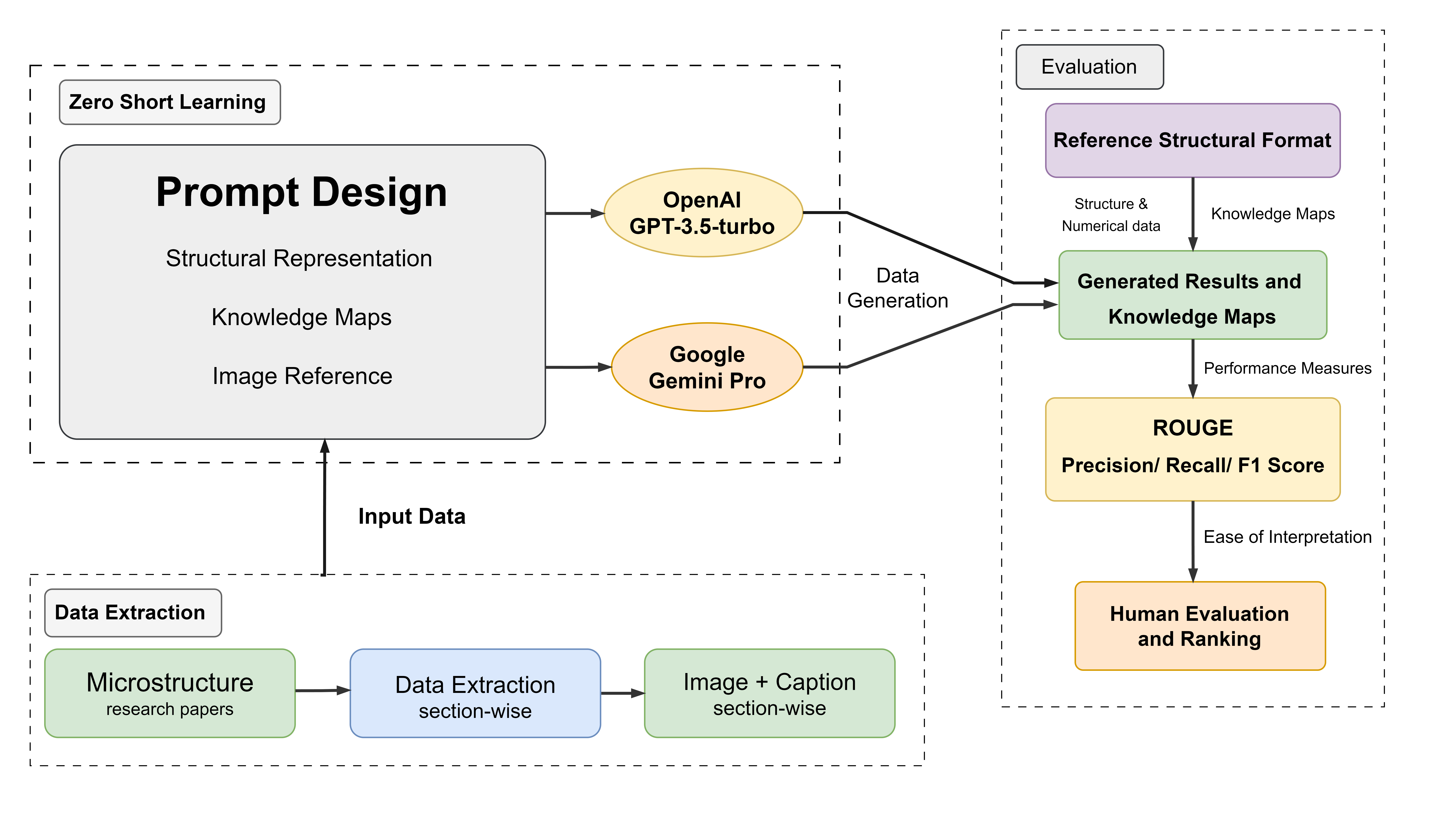}
  \caption{The overall workflow and evaluation framework designed for structural data extraction and the creation of a knowledge map from research papers utilizing zero-shot learning models using GPT-3.5-turbo and GPT-4.}
  \label{fig:final_model_working}
\end{figure}

 In order to generate a Knowledge Graph, the initial step involves prompting the zero-shot LLMs models. See Figure \ref{fig:final_model_working}. This prompting generates a summary of the given text that prioritizes critical processes and key results within the prescribed context, thereby adopting a process--centric approach. This methodology encompasses essential and valuable findings from the paper, including key insights and results obtained through experiments associated with these processes.

\par The subsequent phase involves instructing the model to convert the generated text data from the previous step into a graph representation format for the Knowledge Graph. We utilize the JSON (JavaScript Object Notation) format to store Knowledge Graph data, where nodes, edges, and associated labels are represented. In JSON, nodes are structured as key-value pairs, denoting entities or objects. Edges are connections between these nodes, defining relationships, while labels provide descriptive information about these nodes and edges. This structured JSON format facilitates easy visualization and plotting using any compatible library. Employing LLMs for content extraction yields considerable improvements compared to the previous approaches available in the literature as described in Section \ref{introduction}. This refinement ensures that the resulting knowledge map comprehensively encompasses process details, key insights, and obtained results.


\section{Evalution}
\noindent To assess the structured text within the engineering domain generated by LLM models OpenAI ChatGPT (GPT-3.5-turbo-1106) and Google Gemini Pro, we adopt the Recall-Oriented Understudy for Gisting Evaluation (ROUGE) metric, as detailed by Cohan et al. \cite{cohan2016revisiting}. ROUGE, designed for the evaluation of generated text in tasks such as text generation and machine translation, systematically gauges similarity through precision, recall, and F1 score between the generated text and reference texts.

The performance of the model is assessed using 7 journal articles that comprised both their text and structural formats, in alignment with the research paper's description. The textual data are collected as part of the data collection methods outlined in this paper. The structural data was carefully extracted and verified, and its accuracy is evaluated.

Additionally, in ROUGE evaluation, we employ two measures. The first, Exact Match, checks how precisely the model's text overlaps word-for-word with the reference text, focusing on identical matches. The second, Relaxed Match, considers partial or synonymous matches, making the evaluation more adaptable for assessing content similarity. This dual approach strengthens the reliability of evaluation metrics in natural language processing research, offering a thorough assessment of how well the generated text performs compared to reference standards. By utilizing ROUGE metrics like Precision, Recall, and F1-score, we ensure a standardized and quantitative assessment of text similarity, giving insights into both faithfulness to the reference text of the article and the model's ability to express concepts flexibly. To mitigate the influence of discontinuities in sentences (e.g. due to periods, comas, semicolons), we eliminate such discontinuities from both ground standard and LLMs generated outputs during evaluation.

Since the evaluation of the Knowledge Graph cannot be created in an automated manner based on some metric, when ground truth data is not available, we need to utilize qualitative principles to evaluate the results. The evaluation can be performed by manual inspection of the knowledge graph from the given text considering some criterion and ranking for it. Figure \ref{fig:final_model_working} serves as a visual guide, presenting an overview of the research methodology and the evaluation framework employed in this study.

\section{Analyses and discussion}
\noindent The result shown in Figure \ref{fig:prompt_engineering_working} showcasing the outcomes of experiments detailed in Section 2 of our example input paper \cite{TARZIMOGHADAM2015291}, encompasses a comparison of structurally formatted data extracted from the text of the research article using prompt--based ChatGPT models. Additionally, it includes the generation of knowledge maps from both approaches. Moreover, Figure \ref{fig:final_model_working} illustrates the overall workflow and evaluation process, incorporating the comparison of zero-shot learning against a structural format, the performance contrast between GPT-3.5-turbo-1106 and Google Gemini Pro, and the assessment of knowledge graphs. This comprehensive depiction provides a thorough understanding of the methodologies employed and their resulting outcomes.

\subsection{Structural Representation from the text}

\noindent The performance of prompt-based ChatGPT models is evaluated using the GPT-3.5-turbo-16k model. These models are designed for tasks involving generating and matching text, with a specific focus on extracting structured information from given text.

To assess the effectiveness of the models, we employed two metrics: exact match accuracy and relaxed match performance, utilizing the ROUGE evaluation method. ROGUE provides valuable insights into how well the machine generated text aligns with the reference, both in exactness and inclusiveness. The formulas for both exact match and relaxed match, including the F1 score using ROUGE metrics are given by \cite{cohan2016revisiting}

\paragraph{Exact Match}

\begin{equation}
Precision = \frac{\text{ Number of exact matches}}{\text{Total number of items in generated text}}
\end{equation}

\begin{equation}
Recall = \frac{\text{Number of exact matches }}{\text{Total number of items in reference text}}
\end{equation}

\begin{equation}
\text{F1 Score} = \frac{\text{2 $\times$ Precision $\times$ Recall}}{\text{Precision + Recall}}
\end{equation}

\paragraph{Relaxed Match}

\begin{equation}
Precision = \frac{\text{ Number of exact matches (allowing variations)}}{\text{Total number of items in generated text}}
\end{equation}

\begin{equation}
Recall = \frac{\text{Number of exact matches (allowing variations)}}{\text{Total number of items in reference text}}
\end{equation}

\begin{equation}
\text{F1 Score} = \frac{\text{2 $\times$ Precision $\times$ Recall}}{\text{Precision + Recall}}
\end{equation}

\noindent These formulas provide a concise representation of the evaluation process using ROUGE metrics, including the F1 score that balances precision and recall.

The accuracy of exact matches was assessed using the average values of ROUGE-1 and ROUGE-2, as described in \cite{cohan2016revisiting}. While ChatGPT models demonstrate a lower accuracy with an F1 score of 0.40 in exact match comparisons, this highlights the challenge of achieving precise text reproduction, as indicated by slight variations from the reference text.

Remarkably, the ChatGPT models exhibit exceptional performance in relaxed matching, as assessed by average ROUGE-L and ROUGE-Lsum scores \cite{cohan2016revisiting}. ROUGE-L computes the longest common subsequence for the entire text, ignoring newlines, while ROUGE-Lsum calculates the longest common subsequence for each pair of sentences and aggregates these scores over the entire summary. Table \ref{tab:performance_evaluation} shows a comparison of structurally formatted data extracted from research paper text using prompt-based ChatGPT models, along with the generation of knowledge maps from both approaches. As reported in Table \ref{tab:performance_evaluation} they achieved an impressive F1 score of 0.479 for the structural data format, emphasizing their ability to capture the overall content and structure of the text. This score, derived from ROUGE-1, ROUGE-2, and ROUGE-L measurements, underscores the models' proficiency in comprehending the essence of the information, even when expressed with subtle differences. A similar comparison of performance based on Google Gemini is shown in Table \ref{tab:performance_evaluation2}. A comparison of content extracted by OpenAI ChatGPT and Google Gemini is shown in the \ref{appendixA}.



\begin{table}[h]
    \centering
    \caption{Performance Evaluation of ChatGPT (GPT-3.5-turbo-1106)}
    \label{tab:performance_evaluation}
    \begin{tabularx}{\textwidth}{Xccccccccc}
        \toprule
        & \multicolumn{3}{c}{\textbf{Exact Match}} & \multicolumn{3}{c}{\textbf{Relaxed Match}} \\
        \cmidrule(r){2-4} \cmidrule(r){5-7}
        \textbf{Pair} & \textbf{Recall} & \textbf{Precision} & \textbf{F1 score} & \textbf{Recall} & \textbf{Precision} & \textbf{F1 score} \\
        \midrule
        DataSet1 & 0.47872 & 0.35537 & 0.31159 & 0.34042 & 0.45714 & 0.39024 \\
        DataSet2 & 0.64601 & 0.36111 & 0.39097 & 0.46902 & 0.42741 & 0.44725 \\
        DataSet3 & 0.58333 & 0.527559 & 0.49999 & 0.54166 & 0.67532 & 0.601156 \\
        \midrule
        Average & 0.569358 & 0.414681 & 0.400857 & 0.450373 & 0.519962 & 0.479552 \\
        \bottomrule
    \end{tabularx}
\end{table} 

\begin{table}
    \centering
    \caption{Performance Evaluation of Google Gemini Pro}
    \label{tab:performance_evaluation2}
    \begin{tabularx}{\textwidth}{Xccccccccc}
        \toprule
        & \multicolumn{3}{c}{\textbf{Exact Match}} & \multicolumn{3}{c}{\textbf{Relaxed Match}} \\
        \cmidrule(r){2-4} \cmidrule(r){5-7}
        \textbf{Pair} & \textbf{Recall} & \textbf{Precision} & \textbf{F1 score} & \textbf{Recall} & \textbf{Precision} & \textbf{F1 score} \\
        \midrule
        DataSet1 & 0.54910 & 0.61154 & 0.57454 & 0.66045 & 0.74079 & 0.69204 \\
        DataSet2 & 0.48335 & 0.41384 & 0.42998 & 0.67043 & 0.57906 & 0.59991 \\
        DataSet3 & 0.56870 & 0.52056 & 0.51376 & 0.69053 & 0.62657 & 0.61903 \\
        \midrule
        Average & 0.53372 & 0.51531 & 0.50609 & 0.67381 & 0.64880 & 0.63699 \\
        \bottomrule
    \end{tabularx}
\end{table}

\subsection{Knowledge Graph}


\noindent In assessing a Knowledge Graph where ground truth data is unavailable for automated metric--based evaluation, it becomes evident that relying solely on automated numeric metrics is inadequate. Instead, emphasizing qualitative principles becomes important for accurate assessments. By leveraging insights from domain expertise, specific principles are collated that lead to a valuable methodology for evaluating the effectiveness and precision of the Knowledge Graph, particularly in the absence of concrete ground truth data. The created Knowledge Graphs demonstrate the following advantages:
\begin{enumerate}
 \item Clarity of Experimental Process: Clearly outline experimental procedures used to study material microstructures.

\item Showcase Key Findings: Highlight and emphasize key discoveries and important results derived from experiments.

\item Microstructure Context: Provide detailed context about material microstructures, including phases, defects, and interfaces.

\item Connect Microstructural Elements: Establish strong connections between different components of material microstructures.

\item Diverse Data Sources: Gather information from various sources - experiments, simulations, and theories - for a comprehensive view.

\item Clarity in Terminology: Ensure clarity and  terms are free of human-bias.

\item Structured Understanding: Mapping for a clear understanding of relationships among microstructural elements as described in the text.

\item Insightful Representation: Enable easy extraction of critical material insights from the knowledge graph.


\end{enumerate}

\begin{figure}
  \includegraphics[width=1\linewidth]{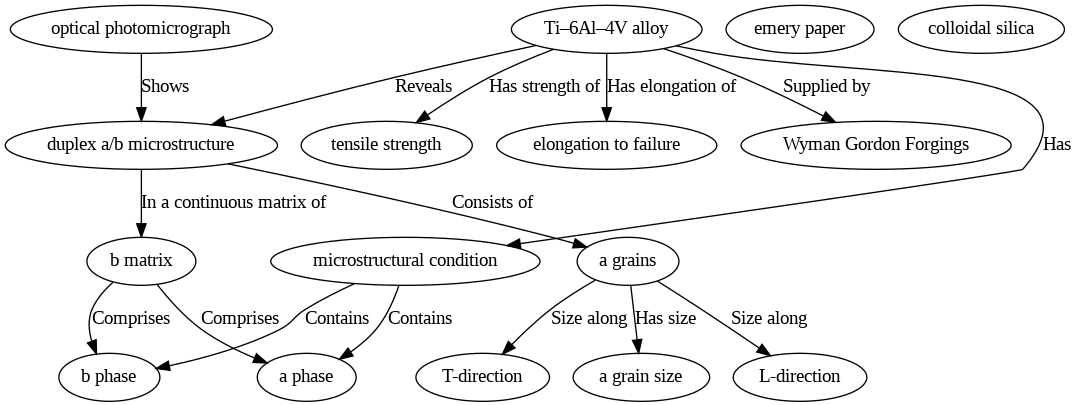}
  \caption{(Generation 1) Creating a knowledge map using a simple LLMs--based relation extraction method with ChatGPT(GPT-3.5-turbo-1106) LLM Model. While the model captures entity relations, it lacks the inclusion of process details, key findings, and crucial results, making the graph less interpretable for analysis.}

  \label{fig:generation_1}
\end{figure}

\begin{figure}
  \includegraphics[width=\textwidth]{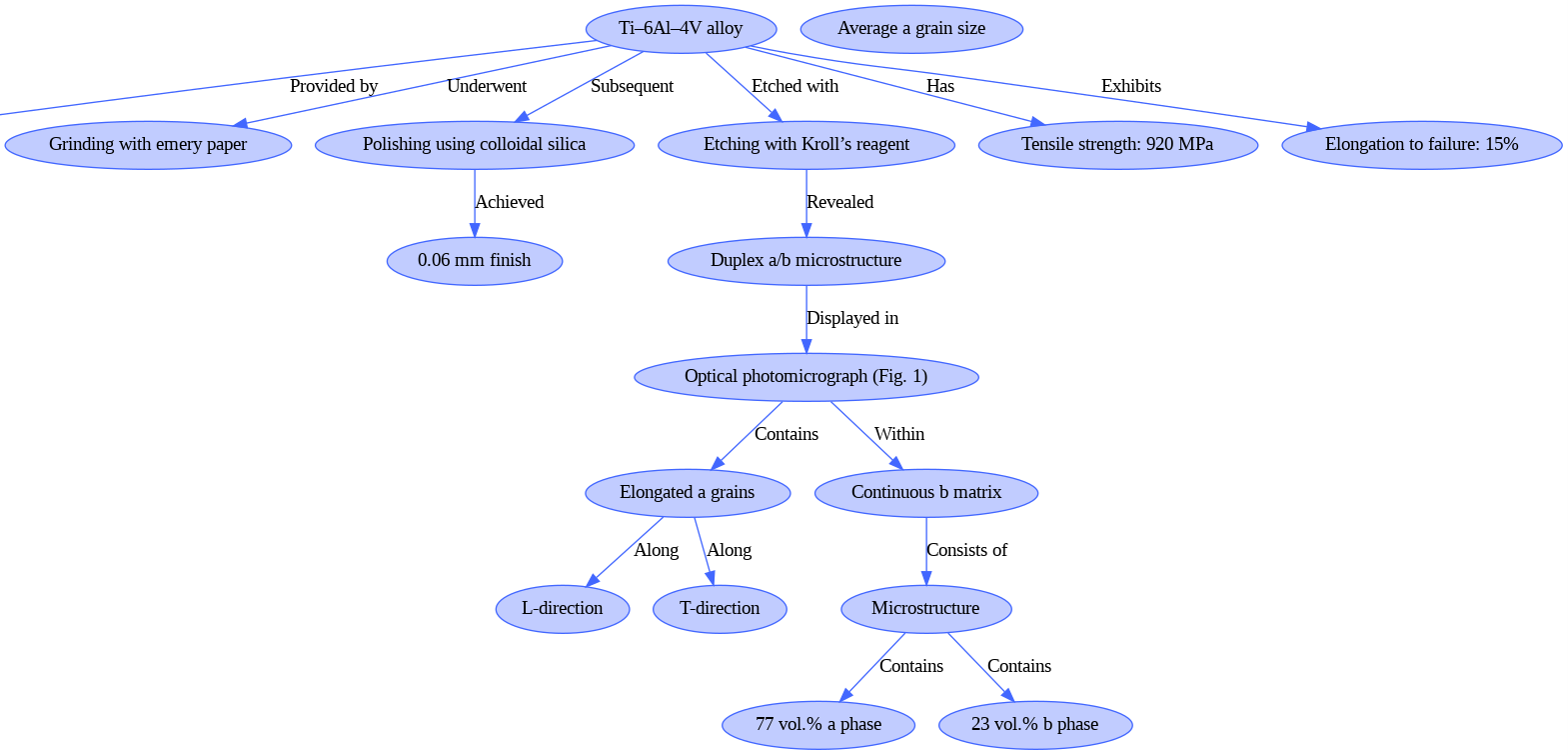}
\caption{(Generation 2) Employing LLMs summary-based approach involves the initial processing of text data through one zero-shot model to generate summaries emphasizing key details. These summaries, crafted via another zero-shot model, are converted into a knowledge graph format, aiding in extracting crucial insights from the input data.}
  \label{fig:generation_2}
\end{figure}

According to these principles, in our use case, we manually inspected the Knowledge Graphs generated with the proposed methods—specifically, Generation 1 (see Figure \ref{fig:generation_1}) and Generation 2 (see Figure \ref{fig:generation_2}). The Generation 1  approach is a simple relation-extraction method (Section \ref{subsec:relationship_extraction}), while Generation 2  is a summary-based approach (Section \ref{subsec:summary_relationship_extraction}). We can conclude that, in our evaluation, the Generation 2 summary-based ChatGPT approach creates a Knowledge Graph of greater quality compared to Generation 1 and aligns with insights gained from domain expertise and our defined principles.

\section{Conclusions}
\noindent This research work introduces a novel NLP  approach using zero-short prompt-based LLMs that address the challenge of structuring unstructured data, specifically in the context of Material Science and Engineering. Using this method, a structured format for data presentation is developed, focusing on extracting pertinent numeric data and key findings from research paper texts. Furthermore, an automated process to generate Knowledge Graphs directly from the research paper text is devised.

This methodology underscores the significant potential in transforming raw and unstructured information into meaningful and valuable insights. Not only text based knowledge but also images are collected along with their corresponding captions from specific sections. These visual aids serve as valuable resources, facilitating the utilization of both the structured data and knowledge maps for easier comprehension and analysis. By employing automated processes, researchers can efficiently extract essential information from complex texts, facilitating a quicker comprehension of the research article's content. This advancement is particularly advantageous for rapid overviews of lengthy research papers, especially when data collection may be of interest.

The results of this research strongly indicate that the combination of NLP techniques, structured data generation, and Knowledge Graph construction holds promise in enhancing the accessibility and usability of scholarly content. As the volume of research literature continues to grow, this approach provides a practical solution to the issue of information overload, enabling researchers to glean swift insights and make well--informed and unbiased decisions without investing excessive time in comprehending intricate texts. By bridging the gap between unstructured data and efficient knowledge extraction, the present methodology contributes to the advancement of research dissemination and comprehension in the field of Material Science and Engineering.

Looking ahead, the future research has a few directions. Initially, the aim is to establish a structured method for evaluating the quality of the generated knowledge graphs from unstructured text. This will help better understand the quality of these graphs a common methodology to judge their overall quality. Additionally, the vision is to extend this approach to different domains and a generalized model to do it. This involves connecting the created knowledge bases and using graphs to predict connections that might be missing between ideas and links in that particular field.

\section*{Conflict of Interest}
\noindent The authors have no relevant financial or non-financial interests to disclose.

\section*{Data availability}
\noindent The raw/processed data required to reproduce these findings cannot be shared at this time as the data also forms part of an ongoing study. Data however can be made available on reasonable request. The tool can be tested at \url{padarthproject.org}.

\bibliographystyle{unsrtnat}
\bibliography{bibliography}

\appendix
\section{An example comparison between OpenAI ChatGPT vs Google Gemini}
\label{appendixA}
\subsection*{\textbf{Given Input}}
\noindent The Ti-6Al-4V alloy that was used in this study was supplied by Wyman Gordon Forgings, Houston, TX. The alloy was ground with emery paper and then polished with the help of colloidal silica to a 0.06 mm finish. After that it was etched in Kroll’s reagent for about 30 s to reveal the duplex a:b microstructure. A triplicate optical photomicrograph of the forged alloy is presented in Fig. 1. This shows the elongated a grains (light phase) in a continuous matrix of $\beta$. The average a grain size along the L-direction is 23.4 $\pm$ 1.0 $\mu$m, while the corresponding a grain size along the T-direction is 16.7 $\pm$ 0.7 $\mu$m. The amounts of a and b phases in this microstructural condition are 77 and 23 vol.
\subsection*{\textbf{Google Gemini Model}}

\subsubsection*{Table: Material Properties}

| Property | Value |
|---|---|
| Material | Ti – 6Al – 4V Alloy |
| Supplier | Wyman Gordon Forgings, Houston, TX | | Surface Finish | 0.06 mm |
| Etching Reagent | Kroll's Reagent |
| Etching Duration | 30 seconds |
| Microstructure | Duplex a:b |
| a Grain Size (L-direction) | 23.4 ± 1.0 $\mu$m |
| a Grain Size (T-direction) | 16.7 ± 0.7 $\mu$m |
| a Phase Volume Fraction | 77
| b Phase Volume Fraction | 23
| Tensile Strength | 920 MPa |
| Elongation to Failure | 15

\subsubsection*{Table: Microstructural Characterization}
| Observation | Description |
|---|---|
| Duplex Microstructure | Elongated a grains in a continuous matrix of b phase | | a Grains (L-direction) | Average size of 23.4 ± 1.0 $\mu$m |
| a Grains (T-direction) | Average size of 16.7 ± 0.7 $\mu$m |
| a Phase Volume Fraction | 77
| b Phase Volume Fraction | 23
\subsection*{\textbf{ChatGPT3.5}}
\subsubsection*{Table: Material Properties}
| Grain Size (L-direction) | 23.4 ± 1.0 $\mu$m | | Grain Size (T-direction) | 16.7 ± 0.7 $\mu$m | | Volume Fraction of a Phase | 77
| Volume Fraction of b Phase | 23
| Tensile Strength | 920 MPa |
| Elongation to Failure | 15

\end{document}